\documentclass[useAMS,usenatbib]{mn2e}
\usepackage{psfrag,graphicx,amsmath}

\newcommand       \be           {\begin{equation}}
\newcommand       \ee           {\end{equation}}
\newcommand       \ba           {\begin{eqnarray}}
\newcommand       \ea           {\end{eqnarray}}
\newcommand       \grad         {\nabla}
\newcommand		  \lan		    {\langle}
\newcommand       \ran          {\rangle}
\newcommand       \apj          {ApJ}

\newcommand       \apjs         {ApJS}
\newcommand       \aap          {A\&A}

\newcommand       \mnras        {MNRAS}

\def\rad{\rm \ rad \, {\rm m^{-2}}}
\def\lesssim{\mathrel{\hbox{\rlap{\hbox{\lower4pt\hbox{$\sim$}}}\hbox{$<$}}}}
\def\gtrsim{\mathrel{\hbox{\rlap{\hbox{\lower4pt\hbox{$\sim$}}}\hbox{$>$}}}}

\title[Spherical Accretion with Anisotropic Thermal Conduction]
{Spherical Accretion with Anisotropic Thermal Conduction}
\author[P. Sharma, E. Quataert, J.~M. Stone]{P. Sharma$^{1}$\thanks{E-mail:
psharma@astro.berkeley.edu}, E. Quataert$^{1}$, J.~M. Stone$^{2}$\\ 
$^{1}$Astronomy Department and Theoretical Astrophysics Center, 
601 Campbell Hall, University of California, Berkeley CA, 94720, USA\\
$^{2}$Department of Astrophysical Sciences, Peyton Hall,
Ivy Lane, Princeton, NJ, 08544, USA}
\begin{document}

\date{Accepted . Received ; in original form }

\maketitle

\label{firstpage}

\begin{abstract}

We study the effects of anisotropic thermal conduction on magnetized
spherical accretion flows using global axisymmetric MHD simulations.
In low collisionality plasmas, the Bondi spherical accretion solution
is unstable to the magnetothermal instability (MTI).  The MTI grows
rapidly at large radii where the inflow is subsonic.  For a weak
initial field, the MTI saturates by creating a primarily radial
magnetic field, i.e., by aligning the field lines with the background
temperature gradient.  The saturation is quasilinear in the sense that
the magnetic field is amplified by a factor of $\sim 10-30$
independent of the initial field strength (for weak fields). In the
saturated state, the conductive heat flux is much larger than the
convective heat flux, and is comparable to the field-free (Spitzer)
value (since the field lines are largely radial).  The MTI by itself
does not appreciably change the accretion rate $\dot M$ relative to
the Bondi rate $\dot M_B$.  However, the radial field lines created by
the MTI are amplified by flux freezing as the plasma flows in to small
radii. Oppositely directed field lines are brought together by the
converging inflow, leading to significant resistive heating.  When the
magnetic energy density is comparable to the gravitational potential
energy density, the plasma is heated to roughly the virial
temperature; the mean inflow is highly subsonic; most of the energy
released by accretion is transported to large radii by thermal
conduction; and the accretion rate $\dot M \ll \dot M_B$.  The
predominantly radial magnetic field created by the MTI at large radii
in spherical accretion flows may account for the stable Faraday
rotation measure towards Sgr A* in the Galactic Center.

\end{abstract}


\section{Introduction}

In magnetized plasmas with a collisional mean free path much greater
than the Larmor radius, there is appreciable heat and momentum
transport along magnetic field lines but negligible transport across
the field lines \citep[][]{bra65}.  \citet{bal00} found that
anisotropic thermal conduction fundamentally alters the convective
stability of dilute stratified plasmas. For a highly collisional
fluid, convection sets in when the entropy increases in the direction
of gravity. A low collisionality plasma, however, is unstable to
convective-like motions regardless of the sign of the temperature
gradient \citep{bal00,qua08}.  For a temperature increasing in the
direction of gravity, horizontal fields are the most unstable (the
magnetothermal instability; MTI), while when the temperature decreases
in the direction of gravity, vertical fields are the most unstable
(the heat flux driven buoyancy instability; HBI).  The MTI \& HBI have
been studied using local simulations by \citet{par05,par07} and
\citet{par08}, respectively.

In this paper, we study the effects of anisotropic thermal conduction
and the MTI on the global dynamics of spherical accretion flows using
axisymmetric magnetohydrodynamic (MHD) simulations.  Thermal
conduction is particularly important for hot low density plasmas in
which the electron mean free path is an appreciable fraction of the
size of the system, as in, e.g., the accretion flow onto Sgr A* in our
Galactic Center and accretion flows onto massive black holes in
elliptical galaxies \citep{tan06}.

There are several motivations for studying anisotropic conduction in
spherical accretion flows.  First, the MTI is predicted to be present
at large radii in spherical accretion flows, for $r \sim r_B$ (where
$r_B \simeq GM/c_s^2$ is the gravitational sphere of influence of the
central object and $c_s$ is the isothermal sound speed of the ambient 
gas).  At
these radii, the inflow is subsonic and there is sufficient time for
the instability to grow \citep{bal00}.  At smaller radii ($r \ll
r_B$), the MTI is unlikely to grow significantly because the inflow is
supersonic (at least according to the classic \citealt{bondi}
solution).  However, if there is resistive dissipation of the magnetic
energy at small radii, conduction may nonetheless be dynamically
important because it provides an efficient way to transport energy to
larger radii, which can suppress the accretion rate onto the central
object \citep{joh06}. As we discuss in more detail later in the paper,
reduction of accretion rate does not specifically require {anisotropic} heat
transport, but is in fact analogous to the suppression of accretion by
magnetic dissipation
seen in the resistive MHD simulations of \citet{igu02}.  

A direct observational motivation for studying the MTI in spherical
accretion flows is provided by mm observations of Sgr A*.  These find
a rotation measure (RM) of $\simeq - 6 \times 10^5$ rad m$^{-2}$ that
has been roughly constant since it was first measured over $\sim 8$
years ago \citep{ait00,bow05,mar07}.  The stability of the observed RM
is difficult to explain if the RM is produced by the accretion flow at
small radii \citep{sha07b}.  In local simulations of the MTI, the
field is amplified by factors of $\sim 10-30$ and becomes aligned with
the temperature gradient \citep{par07}. Thus the MTI at large radii
in spherical accretion flows is expected to produce a primarily radial
magnetic field that could help explain the stable measured RM towards
Sgr A* (a possibility we speculated on in \citealt{sha07b}).

The remainder of this paper is organized as follows.  We describe our
numerical set up in \S 2. Because it is not computationally feasible
to directly simulate both the small radii close to the central object
where most of the gravitationally potential energy is released and the
radii $\sim r_B$ where the MTI develops, we carry out two different
sets of simulations in this paper, covering different radial scales in
the accretion flow.  In \S 3 we present the results of simulations at
$r \sim r_B$ that study the development of the MTI for weak initial
magnetic fields (see Table \ref{tab:tab1}). In \S 4 we describe
simulations at $r \ll r_B$ that study the effect of thermal conduction
and magnetic dissipation on spherical accretion for reasonably strong
initial magnetic fields (see Table \ref{tab:tab2}).  We then conclude
and discuss the astrophysical implications of our results (\S 5).
Throughout this paper we consider the simplified problem of
purely spherical accretion; in future studies, we will examine the
effects of anisotropic thermal conduction and the MTI in rotating
accretion flows.

\section{Numerical Simulations}
We use the axisymmetric ZEUS-2D MHD code in spherical geometry
\citep[][]{sto92a,sto92b}.  We solve the MHD equations with
anisotropic thermal conduction: \ba && \frac{\partial \rho}{\partial
t} + {\bf \grad \cdot} \left ( \rho {\bf V} \right ) = 0, \\ &&
\frac{\partial {\bf V}}{\partial t} + {\bf V \cdot \grad V} =
-\frac{\grad (p + \Pi) }{\rho} + \frac{{\bf J \times B}}{\rho} - \grad \Phi, \\
&& \frac{\partial {\bf B}}{\partial t} = {\bf \grad \times (V \times B
- \eta J)}, \\ && \frac{\partial e}{\partial t} + \grad \cdot (e {\bf
V}) = -(p+\Pi) {\bf \grad \cdot V} - {\bf \grad \cdot Q} + \eta J^2, \\ &&
{\bf Q} = -\chi {\bf \hat{b} (\hat{b} \cdot \grad)} T, \ea where
$\rho$ is the mass density, ${\bf V}$ is the fluid velocity, $p$ is
the pressure, $\Pi$ is an explicit artificial viscosity \citep[e.g., see][]{sto92a}, 
${\bf J}=c({\bf \grad \times B})/4\pi$ is the current
density, $\Phi=-GM/(r-r_g)$ is the gravitational potential due to a
central mass $M$ (we use the pseudo-Newtonian potential of
\citealt{pac80}), $r_g = 2GM/c^2$, ${\bf B}$ is the magnetic field
strength, $\eta$ is the explicit resistivity used in some of our
simulations, $e=p/(\gamma-1)$ is the internal energy density ($\gamma$
is the adiabatic index), ${\bf Q}$ is the heat flux along field lines,
$\chi$ is the thermal diffusivity, ${\bf \hat{b}}$ is the unit vector
along ${\bf B}$, and $T$ is the temperature. For convenience we often
use $\kappa=\chi T/p$ which has the dimensions of a diffusion
coefficient (cm$^2$ s$^{-1}$).  Anisotropic thermal conduction is
added in an operator split fashion with subcycling (e.g., see
\citealt{par05}) and is implemented using the method based on limiters
that guarantees that heat flows from hot to cold regions
\citep[see][]{sha07c}.

We initialize our simulations using the hydrodynamic Bondi solution
with a constant vertical magnetic field $B_z$. Note that the initial
density and temperature gradients are in the radial direction; thus,
the initial field is not everywhere perpendicular to the initial
temperature gradient, unlike in previous local studies of the MTI.  We
consider both weak and strong fields to test the effect of the
strength of the field on the flow dynamics.  The field strength is
specified by the dimensionless number $B_0$, defined by $B_0 = B_z/(8
\pi G M \rho_{\rm out}/r_{\rm out})^{1/2}$, where $r_{\rm out}$ is the
outer radius of the computational domain and $\rho_{\rm out} =
\rho(r_{\rm out})$.  This is equivalent to $\beta(r_{\rm out}) =
B^{-2}_0 [c_s^2/GM/r]|_{r_{\rm out}}$, where $\beta$ is the ratio of
gas pressure to magnetic pressure.

We use an adiabatic index $\gamma=1.5$ so that the initial solution
has a sonic point that lies within the computational domain for our
simulations at $r \sim r_B$ (Table \ref{tab:tab1}).\footnote{For
$\gamma=5/3$ the sonic point lies very close to the origin.  It is,
however, numerically desirable to have the sonic point within the
computational domain so that the inner boundary conditions do not
influence the solution. We have verified that we get similar results 
for a run with $\gamma=5/3$, but with rest of the parameters same as 
the fiducial run R1.}
For our simulations at large radii, we use
the Bondi radius, $r_B= GM/c^2_s$, to normalize length scales in the
problem, while for the simulations at small radii, we use the
gravitational radius of the central object, $r_g = 2GM/c^2$.  We
express time in units of inverse rotation frequency at the outer
radius of the domain, $\Omega_{\rm out}^{-1} = (r_{\rm
out}^3/GM)^{1/2}$.  

The simulations at $\sim r_B$ described in \S 3 (Table \ref{tab:tab1})
go from $r_{\rm in}= 0.04 r_B = 2 \times 10^4 r_g$ to $r_{\rm out}= 16
r_B = 8 \times 10^6 r_g$.  For these simulations, we keep the
temperature at the outer radial boundary fixed at its initial value of
$c_s^2/c^2 = 10^{-6}$, i.e., $T_\infty \simeq 10^7$ K.  The
simulations described in \S 4 (Table \ref{tab:tab2}) go from $2 r_g$
to $256 r_g$.  The temperature is fixed at the virial temperature
$kT_{\rm vir}/m_p \equiv GMm_p/r$ at the outer boundary (except for
one simulation with a very weak initial magnetic field [S2]; in this
simulation there is no significant change in the flow properties
relative to the hydrodynamic Bondi solution, so fixing the temperature
to be that of the initial Bondi solution is more physical). For the
simulations at $\sim r_B$ (\S 3), fixing the temperature at the outer
boundary is reasonable because of the large thermal inertia of the
plasma at large radii.  For the simulations from $2 r_g$ to $256 r_g$
described in \S 4, however, the choice of the outer temperature
boundary condition is more subtle and has a nontrivial effect on the
solution.  Because we initialize the flow with a $\gamma = 1.5 < 5/3$
solution, the initial temperature is significantly less than the
virial temperature.  Fixing $T_{\rm out}$ to its initial value is
inappropriate because dissipation leads to a temperature $\sim T_{\rm
vir}$ throughout most of the domain and thus setting $T_{\rm out}$
equal to its initial value leads to an unphysically large temperature
gradient at $r_{\rm out}$.  Setting $dT/dr = 0$ is unphysical because
this sets the heat flux to 0.  An alternative boundary condition that
we considered, setting a constant heat flux across the outer boundary
(corresponding to $dT/dr \approx$ constant), led to a roughly constant
temperature throughout the domain with $T_{\rm out} \simeq T_{\rm in}
\gg T_{\rm vir}$ after several $\Omega^{-1}_{\rm out}$; this drove all
the mass out of the simulation domain.  We thus settled on the
physically motivated outer boundary condition of $T_{\rm out}=T_{\rm
vir}$.  Factor of few variations in the exact value of $T_{\rm out}$
did not significantly change the results of the simulations.
In both sets of simulations,
inflow/outflow boundary conditions are applied at the radial
boundaries.

\begin{table}
\caption{Simulation parameters for runs with $r \sim r_B$ (\S 3) \label{tab:tab1}}
\begin{tabular}{cccc}
\hline
\hline
Name & Resolution & $\alpha_c^\dagger$ & $B_0^\ddagger$ \\
\hline
R1$^a$ & $60 \times 44$ & $0.2$ & $4.5 \times 10^{-4}$ \\
R2 & $60 \times 44$ & $0.2$ & $4.5 \times 10^{-5}$ \\
R3 & $60 \times 44$ & $0.2$ & $1.34$ \\
D1 & $120 \times 88$ & $0.2$ & $4.5 \times 10^{-4}$ \\
D2 & $120 \times 88$ & $0.2$ & $4.5 \times 10^{-5}$ \\
Q1 & $240 \times 176$ & $0.2$ & $4.5 \times 10^{-4}$ \\
MHD1 & $60 \times 44$ & 0 & $4.5 \times 10^{-4}$ \\
MHD2 & $60 \times 44$ & 0 & $1.34$ \\
\hline
\end{tabular}

$r_{\rm in}= 0.04 r_B$, $r_{\rm out}=16 r_B$ \\
No explicit resistivity is used for these simulations. \\
$^a$ This is our fiducial MTI simulation. \\
$^\dagger$ The conductivity is defined by $\kappa(r) \equiv \alpha_c (GMr)^{1/2}$ (see \S \ref{sec:kappa}). \\
$^\ddagger~B_0$ is the dimensionless magnetic field strength, given by $B_0=B_z/(8\pi GM\rho_{\rm out}/r_{\rm out})^{1/2}$.
\end{table}

\begin{table}
\caption{Simulation parameters for runs with $r \ll r_B$ (\S 4) \label{tab:tab2}}
\begin{tabular}{ccccc}
\hline
\hline
Name & Resolution & $\alpha_c^\dagger$ & $B_0^\ddagger$ & $\frac{\dot{M}_{\rm in}}{\dot{M}_B}^\ast$\\
\hline
S1 & $60 \times 44$ & $0.2$ & $0.32$ & $0.01$ \\
S2 & $60 \times 44$ & $0.2$ & $10^{-13}$ & $0.52$ \\
S3 & $60 \times 44$ & $2$ & $0.32$ & $0.009$ \\
S4 & $60 \times 44$ & $20$ & $0.32$ & $0.039$ \\
MHD3 & $60 \times 44$ & 0 & $0.32$ & $0.028$ \\
\hline
\end{tabular}

$r_{\rm in}= 2 r_g$, $r_{\rm out}=256 r_g$ \\
An explicit resistivity is used for these simulations (eq. [\ref{res}]). \\
$^\dagger$ The conductivity is defined by $\kappa(r) \equiv \alpha_c (GMr)^{1/2}$ (see \S \ref{sec:kappa}). \\
$^\ddagger~B_0$ is the dimensionless magnetic field strength, given by $B_0=B_z/(8\pi GM\rho_{\rm out}/r_{\rm out})^{1/2}$. \\
$^\ast$ time averaged accretion rate through $r_{\rm in}$. \\
\end{table}

The radial grid is logarithmically spaced (grid spacing $dr \propto
r$) so that the number of grid points from $r_{\rm in}$ to $(r_{\rm
in}r_{\rm out})^{1/2}$ is roughly the same as the number of grid
points from $(r_{\rm in}r_{\rm out})^{1/2}$ to $r_{\rm out}$.  A
logarithmic grid gives reasonable resolution at all radii; e.g., for
our simulations at small radii ($2-256 \, r_g$), the cell size at the
smallest radii is $\approx 0.2 r_g$.  A uniform polar grid extends
from $\theta=0$ to $\theta=\pi$.  Reflective boundary conditions are
applied at the poles.  Our standard resolution is $60 \times 44$, but
we have also carried out simulations at higher resolution to assess
convergence. Our calculations use a version of ZEUS that
is not parallel, so that even these modest 2D calculations are
time intensive.

We do not use an explicit resistivity for the $r \gtrsim r_B$
simulations in \S 3 (Table \ref{tab:tab1}).  These simulations are
primarily aimed at studying the MTI for weak fields in which case the
magnetic energy is so small that using an explicit resistivity is not
required to correctly capture the dynamics (in addition, we have
verified that using the explicit resistivity below does not change the
results).  However, for $r \ll r_B$ (\S 4; see Table \ref{tab:tab2}),
the gravitational potential energy of the inflowing plasma is
converted into magnetic energy by flux freezing. The inflow brings
together oppositely directed field lines, which leads to magnetic
dissipation and plasma heating.  This energy is then transported to
large radii by thermal conduction.  To capture this dynamics, it is
important to conserve energy as accurately as possible.  Although we
are not using a conservative code, we can capture a reasonable
fraction of the dissipated magnetic energy using an explicit
resistivity of the form \citep{sto01,igu02} \be \eta = \eta_0 dr^2
\frac{|\grad \times B|}{\sqrt{4 \pi \rho}} \label{res} \ee with
$\eta_0=0.15$ (with this choice for $\eta$, we find that total energy
is conserved to better than 30 \% in our simulations at $r \ll
r_B$). The resistive terms in the induction and internal energy
equations are included using the method of \citet{fle00}.

Although our simulations are two-dimensional, the turbulent magnetic
and velocity fields do not decay at late times, because the
anti-dynamo theorem does not apply when a net magnetic flux threads
the simulations domain. In this case the source current for the
magnetic field is outside of the simulation domain \citep{bal98}.

\subsection{Thermal Conductivity}

\label{sec:kappa}

In a collisional plasma, the thermal conductivity is given by $\kappa
\simeq 4.3 \times 10^{9} T^{5/2}/n$ cm$^2$ s$^{-1}$
\citep[][]{spi62}. However, the inner regions of hot accretion flows
are, in many cases, collisionless with the electron mean free path due
to Coulomb collisions larger than the radius (e.g.,
\citealt{tan06,sha07}); the collisional result is thus inapplicable.
In a collisionless plasma, electrons can in principle carry a free
streaming heat flux as large as $Q \sim n k T v_e \sim 40 \rho v_i^3$
($v_{i,e}=[kT/m_{i,e}]^{1/2}$ is the thermal speed;
\citealt{cow77,sny97}). The diffusion coefficient in this case is
$\kappa \sim v_e r \propto r^{1/2}$, assuming a virial temperature
proportional to $r^{-1}$.  Motivated by the application to
collisionless systems such as Sgr A*, we take $\kappa \propto r^{1/2}$
as the thermal conductivity for our simulations and define the
dimensionless conductivity $\alpha_c$ by \be \kappa \equiv \alpha_c (G
M r)^{1/2}. \label{eq:cond} \ee Although the scaling $\kappa \sim v_e
r \propto r^{1/2}$ is well-motivated, the precise value of $\alpha_c$
is difficult to calculate.  It depends on the electron mean free path
in the collisionless limit, which is determined by the rate of pitch
angle scattering by small-scale (high frequency) turbulent
fluctuations.  Motivated by the results of \citet{sha06b} and
\citet{sha07} on pitch angle scattering in collisionless accretion
flows, we take $\alpha_c \simeq 0.2-2$, i.e., $\kappa = (0.2-2)
(GMr)^{1/2}$, as our fiducial diffusion coefficient.  This value is
much smaller than the maximum free-streaming thermal conductivity that
could be obtained if the electrons are virial.  Our choice for
$\kappa(r)$ is such that the large-scale thermal conduction timescale
($t_{\rm cond}= r^2/\kappa$) is comparable to the free-fall timescale
($t_{\rm ff} = [2GM/r^3]^{-1/2}$).  Thermal conduction can thus lead
to a significant modification of the thermal structure of the plasma
(as we shall see).  Since both $t_{\rm cond}$ and $t_{\rm ff}$ scale
as $r^{3/2}$, this is true at all radii.  In addition, with $t_{\rm
cond} \sim t_{\rm ff}$ on large scales, small-scale perturbations are
effectively isothermal and the MTI grows at its maximal rate
\citep{bal00}.  For the MTI simulations at $\sim r_B$, we carried out
simulations with $\alpha_c$ factors of few smaller and larger than our
fiducial choice of $\alpha_c = 0.2$, and found results essentially
identical to those described here.  For the simulations at $r \sim
2-200 r_g$, resistive heating and thermal conduction significantly
modify the dynamical properties of the accretion flow; in this case,
we show solutions for a variety of values of $\alpha_c$ to explicitly
show the dependence on the conductivity (which is still relatively
modest).

As we discuss in \S3, it turns out that a conductivity $\kappa \propto
r^{1/2}$ has the property that $\grad \cdot {\bf Q} \simeq 0$ for the
{\it initial} Bondi solution near $r \sim r_B$. Since the initial
condition is already approximately a steady state solution even in the
presence of conduction (modulo the MTI), including conductivity does
not affect the structure of the plasma.  We thus carried out
simulations at $\sim r_B$ with other choices for $\kappa(r)$ (namely,
$\kappa \propto r^{3/2}$ and $\kappa \propto r^{-1/2}$).  Although we
do not show detailed results for these simulations, in all cases we
found that the MTI rapidly creates radial magnetic field lines and
that conduction along the radial magnetic field lines forces the
density and temperature structure of the plasma to readjust in order
to satisfy $\grad \cdot {\bf Q} \simeq 0$. This can lead to modest
modifications in the density and temperature profiles of the flow, but
in all cases the accretion rate $\dot {M}$ was within a factor of a
few of the Bondi rate.

\section{Simulations at $r \sim r_B$}

\begin{figure}
\begin{center}
\includegraphics[scale=0.42]{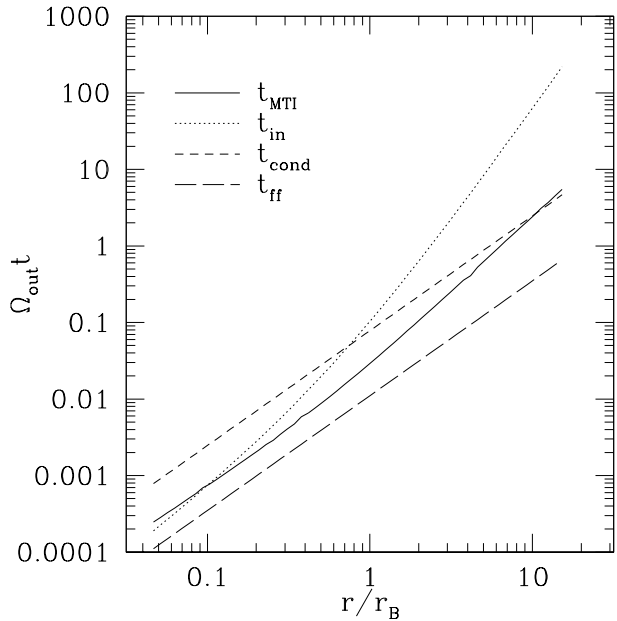}
\caption{Various timescales (in units of $\Omega_{\rm out}^{-1}
=[GM/r_{\rm out}^3]^{-1/2}$) as a function of radius for the initial
Bondi solution: the MTI growth time $t_{\rm MTI}=\gamma_{\rm
MTI}^{-1}$ (the maximum MTI growth rate is given by $\gamma_{\rm
MTI}^2 = [1/\rho][dp/dr][d \ln T/dr]$), $t_{\rm in}=r/V_r$ (the local
infall timescale), $t_{\rm cond}=r^2/\kappa$ (the large scale thermal
diffusion timescale for $\alpha_c = 0.2$), and $t_{\rm
ff}=(r^3/2GM)^{1/2}$ (the local free-fall timescale).  The infall
timescale is much longer than the timescale for the growth of the MTI
for $r \gtrsim r_B$; thus the MTI rapidly modifies the magnetic field
structure.}
\label{fig:fig1}
\end{center}
\end{figure}

Figure \ref{fig:fig1} shows several different timescales
characterizing the initial Bondi solution. For $r \gtrsim r_B$ the
infall timescale ($t_{\rm in}$) is much longer than the timescale for
the MTI to grow ($t_{\rm MTI}$) and the instability grows
exponentially.  The MTI can thus saturate before the plasma flows in.
At smaller radii ($r \lesssim r_B$), however, where the inflow time is
comparable to the MTI growth time, the growth is algebraic and the MTI
is less likely to have a significant effect on the dynamics. Instead,
at $r \ll r_B$, the magnetic field will be primarily amplified by flux
freezing.  Here we discuss our simulations of the MTI at $r \sim r_B$.
In the next section we present the results of simulations at $r \ll
r_B$.

\subsection{The Fiducial MTI Simulation}
\begin{figure}
\begin{center}
\includegraphics[scale=0.42]{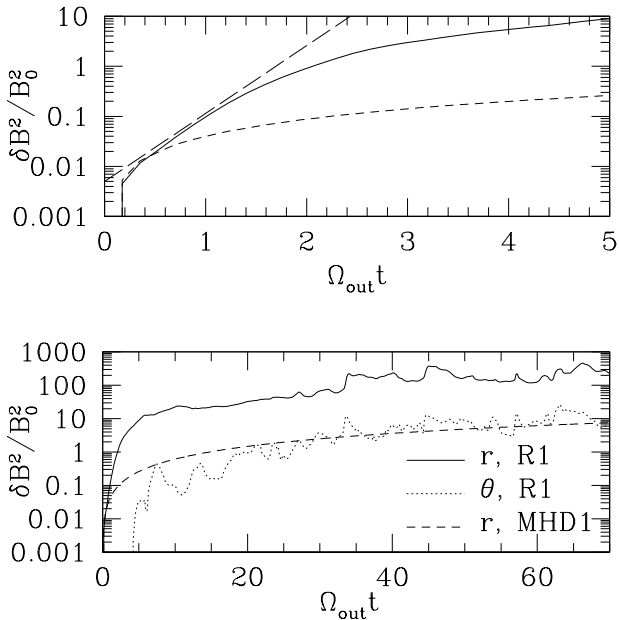}
\caption{Change in volume averaged magnetic energy (radial and $\theta$ 
components normalized to the initial magnetic energy) in a shell from 
$r_{\rm out}/10$ to $r_{\rm out}$ as a function of time.  The top panel shows 
the early-time evolution of
$\delta B_r^2$ (solid line) for the fiducial MTI simulation (R1)
compared to $\delta B_r^2$ for an MHD simulation with the same initial
conditions (MHD1; short-dashed line).  The long-dashed line indicates the
growth rate of the MTI at $r \simeq r_{\rm out}/3$.  The bottom panel
shows the same field components over longer timescales, along with
$\delta B_\theta^2$ (dotted line) for the fiducial run.  $\delta
B^2_{r,\theta} \equiv (\lan B_{r, \theta}^2 \ran_V - B_{r, \theta, 0}^2)$,
where $\lan \ran_V$ denotes volume average and $B_0$ is the
initial field strength.} 
\label{fig:fig2}
\end{center}
\end{figure}

Our fiducial simulation (labeled ``R1'') uses $\kappa(r) = 0.2
(GMr)^{1/2}$ and has an initial magnetic field strength of $B_0=4.5
\times 10^{-4}$, corresponding to $\beta \simeq 10^8$ at $r_{\rm
out}$.  This simulation describes the evolution of the MTI for weak
magnetic fields. We choose the lower resolution simulation R1
as our fiducial case because it finishes relatively quickly (in
$\lesssim$ a week); this enables us to readily study the effects of
different boundary conditions, different choices for the thermal
conductivity, etc. We also present higher resolution calculations
(e.g., D1 \& Q1, which take several months to finish) that are similar
to R1, although with a modest increase in the field amplification by
the MTI (see Fig. [\ref{fig:fig7b}]).

Figure \ref{fig:fig2} shows the $r$ and $\theta$
components of the magnetic energy density as a function of time for
the fiducial simulation and for an MHD simulation without conduction,
but with the same initial conditions.  The energy is averaged from
$r=r_{\rm out}/10$ to $r_{\rm out}$, to isolate the radii where
$t_{\rm in} \gg t_{\rm MTI}$ (see Fig. \ref{fig:fig1}) and the
instability is unaffected by inflow.  The top panel in Figure
\ref{fig:fig2} shows the early linear growth of the MTI, while the
bottom panel shows the longer time evolution.  Because we are
averaging over a range of radii, there is no unique growth rate for
the MTI, but the maximal MTI growth rate evaluated at $r \simeq r_{\rm
out}/3$ provides a reasonable fit to the early time growth
(long-dashed line).  After a few free-fall times (several MTI growth
times; see Fig. \ref{fig:fig1}), the magnetic field ceases to grow
exponentially and the growth becomes algebraic.  This represents the
saturation of the MTI; the later growth of the field is due to flux
freezing.  This can be seen explicitly by noting that the MHD
simulation (short-dashed line) shows the same algebraic growth in the
magnetic field with time, only at a reduced magnitude of ${\bf B}$ because
there is no early MTI growth.  Note that the net increase in magnetic
energy because of the MTI is quite modest: $\beta \gg 1$ even in the
saturated state.

\begin{figure}
\begin{center}
\includegraphics[scale=0.3]{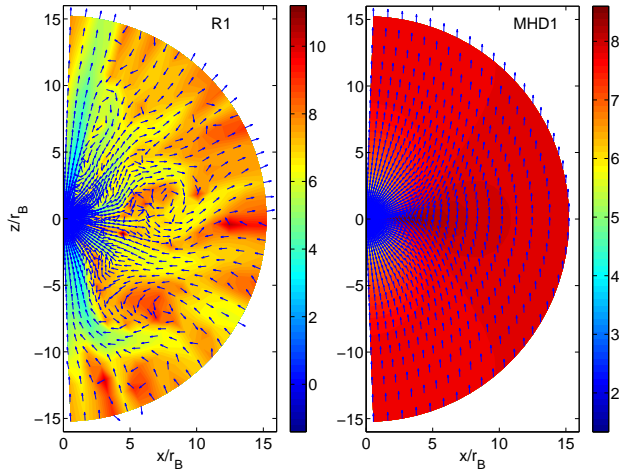}
\caption{${\rm Log}_{10}$ of plasma $\beta~(\equiv 8\pi p/B^2)$ at the
end ($\Omega_{\rm out}t \approx 60$) of the fiducial MTI run R1
(left), and at the end of the MHD simulation MHD1 (right).  Also shown
are arrows indicating the local magnetic field direction. The magnetic
field is turbulent but primarily radial in the presence of anisotropic
conduction. In MHD, the field remains primarily vertical at large
radii, because flux freezing only reorients the field on the (slow)
inflow timescale.}
\label{fig:fig3}
\end{center}
\end{figure}

\begin{figure}
\begin{center}
\includegraphics[scale=0.42]{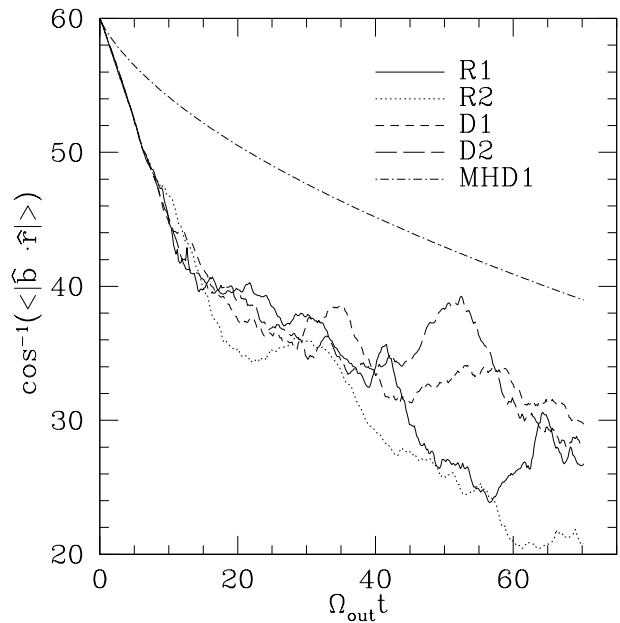}
\caption{Volume averaged angle (in degrees) of the magnetic field lines with
respect to the radial direction as a function of time for different
runs (the average is from $r_{\rm out}/10$ to $r_{\rm out}$, as in
Fig. \ref{fig:fig2}).
}
\label{fig:fig3b}
\end{center}
\end{figure}

Figure \ref{fig:fig3} compares the plasma $\beta$ at the end
($\Omega_{\rm out}t\simeq 60$) of the fiducial run and the
corresponding MHD simulation. The magnetic field unit vectors are also
shown. The plasma $\beta$ with conduction is a factor of $\sim 100$
smaller than without conduction (as is also seen in
Fig. \ref{fig:fig2}), indicating that the MTI alone amplifies the
field by a factor of $\sim 10$.  Since the simulations do not run for
even one infall timescale at $r_{\rm out}$, the magnetic field lines
at large radii do not change significantly in the MHD simulation and
remain primarily vertical.  By contrast, the magnetic field is
completely restructured in the presence of anisotropic conduction
because of the MTI. The magnetic field at the end of the simulation is
primarily radial.  This can also be seen in Figure \ref{fig:fig3b},
which shows the volume averaged angle of the magnetic field with
respect to the radial direction as a function of time for several
runs. In the linear stage of the MTI, the radial field is amplified
exponentially; after saturation, the field direction, although
turbulent, continues to become radial because of flux freezing (as
also occurs in the pure MHD simulation shown in Fig. \ref{fig:fig3b}),
although this occurs on the slower inflow timescale.  Our conclusion
that the MTI generates primarily radial field lines in the saturated
state is fully consistent with the local MTI simulations of
\citet{par05,par07}.

\begin{figure}
\begin{center}
\includegraphics[scale=0.42]{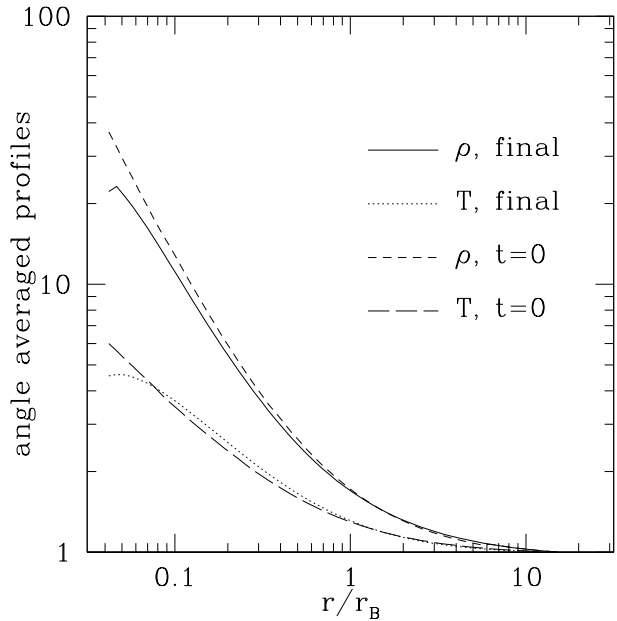}
\caption{Density (short-dashed line) and temperature (long-dashed
line) as a function of radius for the initial Bondi solution; time and
angle averaged density (solid line) and temperature (dotted line) at
late times as a function of radius for the fiducial MTI simulation
R1. Density and temperature are normalized to their values at $r_{\rm
out}$.  There is very little change in the radial density and
temperature profiles during the simulation.}
\label{fig:fig4}
\end{center}
\end{figure}

To understand the saturation of the MTI, it is useful to consider the
linear dynamics of the instability.  In the WKB limit, and assuming
rapid conduction and weak magnetic fields, the linear growth rate is
given by \citep{qua08} \be
\label{eq:growth}
\gamma^2 \approx \frac{1}{\rho}\frac{dp}{dr} \left ( \frac{d \ln
T}{dr} \right ) \left [ (1-2b_r^2) \frac{k_\theta^2}{k^2} + \frac{2
b_\theta b_r k_\theta k_r}{k^2} \right ], \ee where $k_r$ and
$k_\theta$ are the $r$ and $\theta$ wavenumbers, and $b_r$ and
$b_\theta$ are the $r$ and $\theta$ magnetic field unit vectors. Our
initial field is vertical with a $\theta$ dependent $b_r$ and
$b_\theta$.  In the saturated state, where $b_r \simeq 1$ and
$b_\theta \ll 1$, the fastest growth rate of the MTI is $\gamma \simeq
|[1/\rho][dp/dr][d \ln T/dr]|^{1/2} b_\theta$, a factor of $\sim
b_\theta$ smaller than the growth rate in the initial configuration.
Growth occurs only for modes satisfying $k_\theta/k_r \lesssim 2
b_\theta$, i.e., for small radial wavelengths.  The significant
decrease in the growth-rate of the MTI for nearly radial fields
qualitatively explains why magnetic field reorientation leads to
saturation of the instability.  In addition, even if there is growth
at late times, it is on small radial scales and so is unlikely to
modify the large-scale field structure or magnetic energy.  The MTI
can also saturate by making $d\ln T/dr \approx 0$, i.e., by making the
plasma isothermal. Whether the plasma becomes isothermal or not
depends on $\kappa (r)$ (and on the boundary conditions), which
influences how the density and temperature adjust to make ${\bf \grad
\cdot Q} \simeq 0$ in steady state. Figure \ref{fig:fig4} shows that
for our fiducial simulation, there is very little change in the
temperature and density profiles during the simulation.  For our
choice of $\kappa \propto r^{1/2}$, it turns out that the initial
Bondi flow nearly satisfies $\chi r^2 dT/dr = {\rm constant}$ near
$\sim r_B$.  Thus when the MTI reorients the field lines to be
primarily radial, little evolution in $\rho(r)$ and $T(r)$ is required
to make ${\bf \grad \cdot Q} \simeq 0$.  As noted in \S
\ref{sec:kappa}, we carried out simulations with different $\kappa(r)$
in order to assess its effects on the flow dynamics (although we
believe that the fiducial choice considered here is the most physical
for collisionless systems).  In all cases, we found that the MTI
reorients the field lines to be primarily radial and that $\rho(r)$
and $T(r)$ adjust so that ${\bf \grad \cdot Q} \simeq 0$.

\begin{figure}
\begin{center}
\includegraphics[scale=0.42]{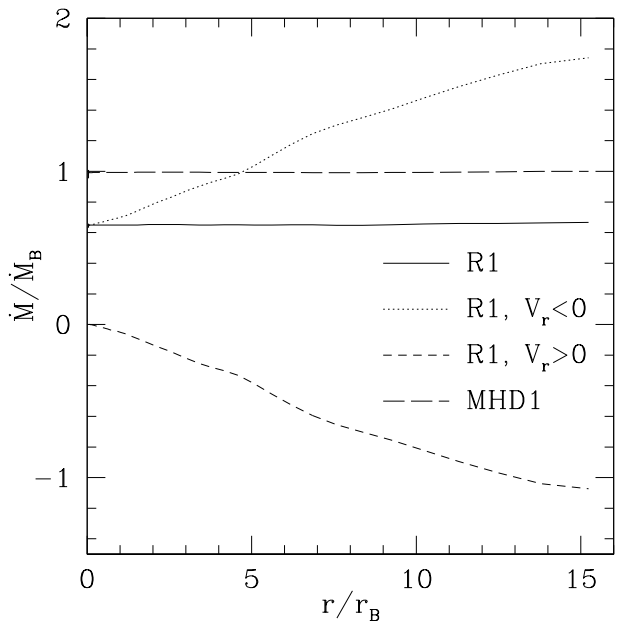}
\caption{Time and angle averaged mass accretion rate ($\dot{M}$) in
the fiducial MTI simulation R1 at late times, as a function of radius
(normalized to the Bondi accretion rate, $\dot{M}_B$): net mass
accretion rate (solid line), mass inflow rate (dotted line; with
$V_r<0$), and mass outflow rate (short-dashed line; with $V_r>0$). The
mass accretion rate in MHD is shown by the long-dashed line.}
\label{fig:fig5}
\end{center}
\end{figure}

Figure \ref{fig:fig5} shows the time and angle averaged mass accretion
rate in the saturated state as a function of radius.  In MHD the
velocity is always radial and inwards and the mass accretion rate is
equal to the Bondi rate (for this $\beta \gg 1$ simulation).  The
solid line in Figure \ref{fig:fig5} shows the net mass accretion rate
for the fiducial run, $\dot{M}=2 \pi r^2 \int \rho V_r \sin \theta d
\theta $; also shown are the mass fluxes of material with $V_r > 0$
and $V_r < 0$.  Although there are MTI-driven convective motions, they
are modest and the net accretion rate is only slightly smaller than
the Bondi accretion rate.  This is partially a consequence of the fact
that the density and temperature profiles change very little in the
course of the simulation (Fig. \ref{fig:fig4}).  However, even for
different choices of $\alpha_c$ and $\kappa(r)$, we still found $\dot
M \sim \dot M_B$ for MTI simulations with initially weak fields.  The
MTI by itself thus does not significantly change the accretion rate
onto the central object.  This is in contrast to the very strong
suppression of $\dot{M}$ found in the next section by the combined
action of magnetic dissipation, plasma heating, and thermal conduction
at $r \ll r_{B}$.

\begin{figure}
\begin{center}
\includegraphics[scale=0.42]{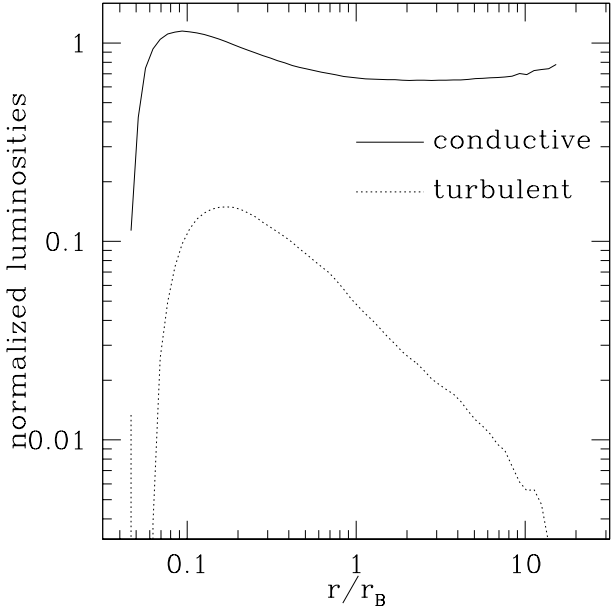}
\caption{Time and angle averaged conductive (solid line; $Q_c$) and
turbulent (dotted line; $Q_t$) luminosities ($4\pi r^2 \times$ energy
fluxes) as a function of radius for the fiducial MTI simulation R1;
see \S3.1 for the definition of $Q_t$.  The luminosities are
normalized to the field free conduction luminosity at $r_{\rm out}$.}
\label{fig:fig6}
\end{center}
\end{figure}

Figure \ref{fig:fig6} shows time and angle averaged luminosities ($4
\pi r^2 \times$ energy fluxes) for energy transport by conduction and
MTI-initiated convection.  The luminosities are normalized to the
field free (``Spitzer'') conductive luminosity at $r_{\rm out}$.  The
radial conductive luminosity ($4 \pi r^2 Q_c$) is roughly constant as
a function of radius for $r \gtrsim 0.1 r_B$, as would be expected in
steady state.  The conductive heat flux is similar to the local
field-free value at all radii because the field lines are primarily
radial. The non-conductive energy transport has contributions from the
kinetic energy flux ($\rho V^2 V_r/2$), the enthalpy flux
($\gamma/(\gamma-1) p V_r$), and the Poynting flux.  The latter is
negligible for the weak magnetic field values considered here.  Both
the kinetic energy flux and the enthalpy flux can be split up into
terms proportional to the mass accretion rate (the advective flux of
energy $Q_a$ due to the bulk motion of the fluid) and terms which
represent the turbulent/convective transport of energy ($Q_t$).
Mathematically, this can be seen by decomposing the fluid variables
into mean and fluctuating components; e.g., the radial velocity is
taken to be $V_r = \lan V_r \ran + \delta V_r$, where $\lan \ran$
represents an appropriate average and $\delta$ represents deviation
from the average.  For our purposes, the average is taken over
$\theta$ at each time at a given $r$. With this decomposition, the
average kinetic energy flux is given by \be \nonumber \frac{1}{2}\lan
\rho V^2 V_r \ran = Q_{K,a} + Q_{K,t}, \ee with \ba Q_{K,a} &=&
\frac{1}{2} \lan \rho V_r \ran [\lan V_r \ran^2 + \lan V_\theta \ran^2
+ \lan \delta V_r^2 \ran + \lan \delta V_\theta^2 \ran ], \\ \nonumber
Q_{K,t} &=& \frac{1}{2} \lan \delta (\rho V_r) (\delta V_r^2 + \delta
V_\theta^2) \ran + \lan \delta (\rho V_r) \delta V_r \ran \lan V_r
\ran \\ &+& \lan \delta (\rho V_r) \delta V_\theta \ran \lan V_\theta
\ran, \ea where $Q_{K,a}$ is the advective flux of kinetic energy and
$Q_{K,t}$ is the turbulent/convective kinetic energy flux.  Similarly,
the average enthalpy flux can be written as \be \nonumber
\frac{\gamma}{\gamma-1} \lan p V_r \ran = Q_{e,a} + Q_{e,t}, \ee with
\ba Q_{e,a} &=& \frac{\gamma}{\gamma-1} \lan \rho V_r \ran \lan c_s^2
\ran , \\ \nonumber Q_{e,t} &=& \frac{\gamma}{\gamma-1} [ \lan V_r
\ran \lan \delta \rho \delta c^2_s \ran + \lan \rho \ran \lan \delta
V_r \delta c^2_s \ran \\ &+& \lan \delta \rho \delta V_r \delta c^2_s
\ran ], \label{eq:enthalpy} \ea where $c_s^2 = p/\rho \simeq kT/m_p$.
Figure \ref{fig:fig6} shows the total turbulent energy flux in the MTI
simulations at large radii, $Q_t = Q_{e,t} + Q_{K,t}$; the turbulent
energy transport -- which is dominated by the $\lan \delta V_r \delta
c_s^2 \ran$ term in equation (\ref{eq:enthalpy}) -- is much smaller than the
conductive energy transport at all radii.

\subsection{Effect of $B_0$}
\label{sec:B0}

If the MTI saturates by making the magnetic field lines primarily
radial, as we showed in the previous section, it is reasonable to
expect that the saturation magnetic field strength will scale with the
initial magnetic field strength (since a smaller radial field is
required to make an initially weaker vertical magnetic field
radial). To test this hypothesis we carried out the simulation R2 with
a weaker vertical magnetic field ($B_0=4.5 \times 10^{-5})$,
corresponding to $\beta \simeq 10^{10}$ at $r_{\rm out}$.  All other
parameters are same as the fiducial run.

\begin{figure}
\begin{center}
\includegraphics[scale=0.42]{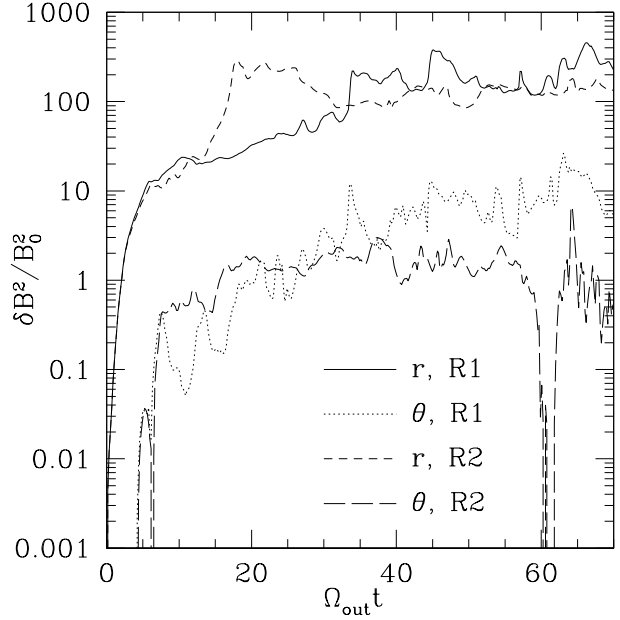}
\caption{Volume averaged magnetic energy (normalized to the initial
magnetic energy) in a shell from $r_{\rm out}/10$ to $r_{\rm out}$ as
a function of time: $\delta B_r^2/B_0^2$ (solid line), $\delta
B_\theta^2/B_0^2$ (dotted line) for the fiducial MTI simulation R1;
$\delta B_r^2/B_0^2$ (short-dashed line), $\delta B_\theta^2/B_0^2$
(long-dashed line) for R2 (which has an initial field strength $B_0$
10 times smaller than R1); $\delta B^2_{r,\theta} = (\lan B_{r,
\theta}^2 \ran_V - B_{r,\theta,0}^2)$, where $\lan \ran_V$ denotes
volume average and $B_0$ is the initial field strength.}
\label{fig:fig7}
\end{center}
\end{figure}

Figure \ref{fig:fig7} shows the results from run R2 compared to those
of R1.  The most interesting result is that even nonlinearly $\delta
B_{r,\theta}^2/B_0^2$ for R1 and R2 are similar.  That is, the late
time magnetic energy is roughly proportional to the initial magnetic
energy in the domain.  This indicates that the saturation of the MTI
is quasilinear: for weak magnetic fields, the MTI saturates when the
field lines become radial, which requires a $\delta B \propto B_0$.

We also carried out simulations with much stronger initial magnetic
fields, $B_0=1.3$, corresponding to $\beta \simeq 10$ at $r_{\rm out}$
(run R3 with conduction and run MHD2 without conduction). The magnetic
field is sufficiently strong that the MTI is suppressed by magnetic
tension.  Indeed, we find that there are only small differences
between the simulations with and without conduction.  In both cases
the magnetic field at small radii is amplified by the converging
inflow. Magnetic tension prevents plasma infall in the equatorial
region; mass accretion is dominated by the polar regions. There is a
strong equator-pole anisotropy because of the strong magnetic
field. The polar regions are cold, low density, and magnetically
dominated ($\beta \ll 1$), with large radial infall velocities.  We
find a modest reduction in the accretion rate for these strong-field
simulations at $\sim r_B$, with the mass accretion rate through
$r_{\rm in}$ being $\simeq 25\%$ of the Bondi rate.  Finally, we find
that including an explicit resistivity does not have a significant
effect on the flow dynamics at large radii even for energetically
significant magnetic fields.  This is in contrast to the simulations
described in \S 4 in which there is significant magnetic dissipation
when $\beta \sim 1$ at $r \ll r_B$.  The difference is probably that
the supersonic inflow at $r \ll r_B$ is much more effective than the
slow inflow velocities at $\sim r_B$ at forcing
reconnection/dissipation of field lines.

\subsection{Effect of Resolution}

\begin{figure}
\begin{center}
\includegraphics[scale=0.42]{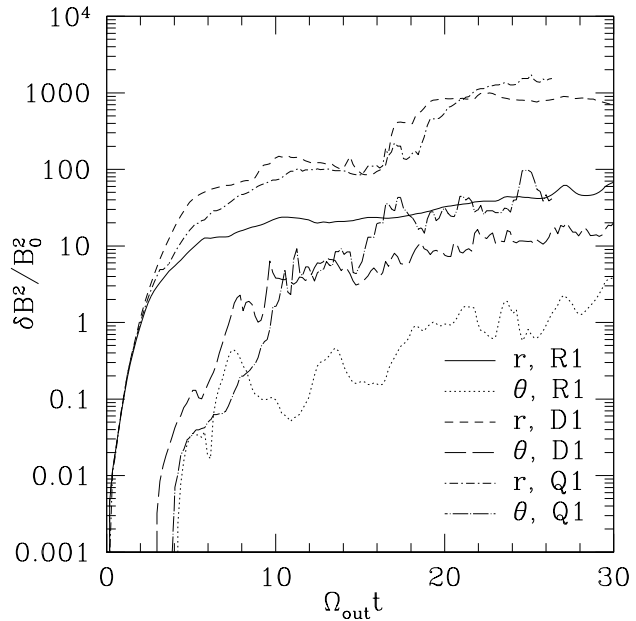}
\caption{Volume averaged magnetic energy (normalized to the initial
magnetic energy) in a shell from $r_{\rm out}/10$ to $r_{\rm out}$ as
a function of time, comparing different resolution runs: $\delta
B_r^2/B_0^2$ (solid line), $\delta B_\theta^2/B_0^2$ (dotted line) for
the fiducial run R1; $\delta B_r^2/B_0^2$ (short-dashed line), $\delta
B_\theta^2/B_0^2$ (long-dashed line) for D1, with twice the resolution
as R1; and $\delta B_r^2/B_0^2$ (short dot-dashed line), $\delta
B_\theta^2/B_0^2$ (long dot-dashed line) for Q1, with four times the
resolution of R1. As before, $\delta B^2_{r,\theta} = (\lan B_{r,
\theta}^2 \ran_V - B_{r,\theta,0}^2)$, where $\lan \ran_V$
denotes volume average and $B_0$ is the initial field strength.}
\label{fig:fig7b}
\end{center}
\end{figure}

To assess the effect of resolution we carried out several higher
resolution runs: run D1, which is analogous to the fiducial simulation
R1 but with twice the resolution, run D2 which is analogous to R2 but
with twice the resolution, and run Q1 which has four times the
resolution of R1.  While the overall dynamics is very similar to the
lower resolution runs, the magnetic energy is a factor of few larger
in the higher resolution simulations.  Figure \ref{fig:fig7b} compares
the fiducial run R1 with D1 and Q1, and shows that both $\delta B_r^2$
and $\delta B_\theta^2$ are larger in higher resolution runs than they
are in R1, by a factor of $\simeq 5$.  The same is true for D2
compared with R2. In particular, Figure \ref{fig:fig7b} shows that the
phase of linear exponential magnetic field amplification occurs for a
longer period of time in the higher resolution simulations.  Since the
late-time magnetic field growth is simply flux freezing of the
early-time field, this leads to an enhanced field at all times.  The
fastest growing MTI modes for approximately radial magnetic fields are
those with $2k_r/k_\theta \gtrsim b_\theta^{-1}$, which is a
resolution dependent criterion.  Higher resolution simulations can
thus resolve growth for longer times and to larger radial fields,
i.e., to smaller $b_\theta$.

Although the fastest growth of the MTI occurs at small scales,
nonlinearly one expects the energy contained in small scales to be
subdominant for sufficiently high resolution simulations.  Indeed,
Figure \ref{fig:fig7b} shows that both the radial and polar magnetic
energies are similar for the higher resolution runs D1 and
Q1. This indicates
that nonlinear saturation of the MTI is reasonably independent of
resolution for sufficiently high resolution simulations (e.g., runs D1
and Q1).

\section{Simulations at $r \ll r_B$}

As shown in the previous section, the MTI generates a radial field at
$\sim r_B$ in dilute spherical accretion flows.  As the plasma flows
in to yet smaller radii, this field will be amplified by flux
freezing.  It is well known that for inflow according to the Bondi
solution \citep{bondi}, the magnetic energy increases more rapidly
with decreasing radius than the gravitational energy of the gas, and
eventually the field becomes energetically important (e.g.,
\citealt{shvartsman1971}).  \citet{igu02} showed that when $\beta \sim
1$ magnetic dissipation of the amplified field becomes significant and
the dynamics of the inflow changes appreciably.  Resistive heating
increases the temperature of the plasma and may drive thermal
convection.  The accretion rate onto the central object decreases by a
few orders of magnitude. In this section, we examine the modification
to these results caused by anisotropic thermal conduction.  Our
general conclusion that mass accretion is significantly suppressed is
consistent with that of \citet{igu02}, but we find that the conductive
transport of energy is more important than the convective transport.

Our simulations at small radii have $r_{\rm in}=2r_g$ and $r_{\rm
out}=256 r_g$; their properties are summarized in Table
\ref{tab:tab2}.  The initial conditions are chosen from the same Bondi
solution used in the previous section.  Run S1 takes $\alpha_c=0.2$
and $B_0 \simeq 0.32$.  This value of $B_0$ corresponds to $\beta
\simeq 1$ at $r_{\rm out}$.  \citet{igu02} have shown that a weaker
field at $r_{\rm out}$ means that significant dissipation does not
begin until smaller radii (and later times), but that otherwise the
results do not depend sensitively on the choice of $B_0$.  In addition
to the fiducial value of $\alpha_c = 0.2$, we also performed
simulations with $\alpha_c = 2$ and $20$; the latter corresponds
roughly to the maximum rate of heat transport by free-streaming
electrons.  We present results for a variety of $\alpha_c$ but focus
on the $\alpha_c = 2$ simulation (S3) because it most clearly
demonstrates the relevant physics.

S2 and MHD3 represent control simulations.  S2 is a simulation with
anisotropic conduction but with a very weak field.  As expected, it is
effectively a laminar hydrodynamic Bondi solution (see
Fig. \ref{fig:vr}), with no evidence for the MTI or dynamically
significant resistive heating.  MHD3 is a resistive MHD simulation
with $\beta \simeq 1$ at $r_{\rm out}$, analogous to the simulations
performed by \citet{igu02}.

\subsection{Results}

\begin{figure}
\begin{center}
\includegraphics[scale=0.42]{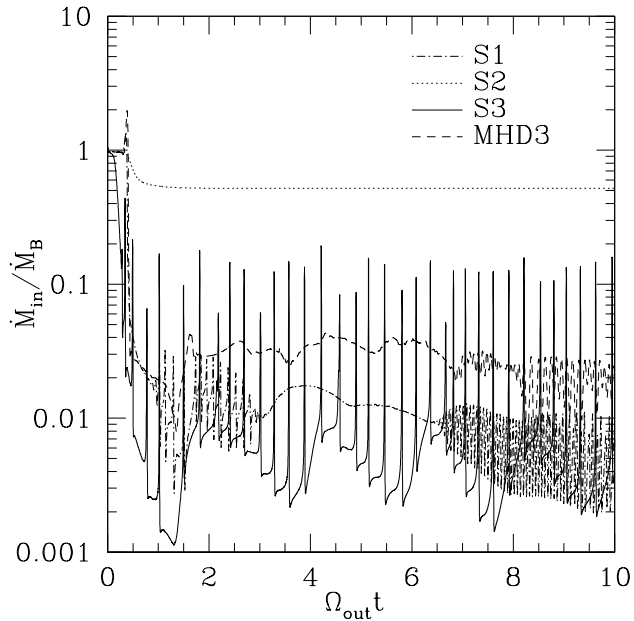}
\caption{Mass accretion rate (normalized to the initial Bondi rate) at
$r_{\rm in} = 2 r_g$ as a function of time for runs S1, S2, S3, \&
MHD3, all listed in Table \ref{tab:tab2}. The mass accretion rate in
the presence of strong fields and resistive dissipation is reduced
significantly relative to the Bondi rate, irrespective of whether
conduction is present. Note that the mass accretion rate for run
S3 shows very regular ``flares;" these persist even if we vary many of
the numerical details of the calculation (see text for details).}
\label{fig:fig8}
\end{center}
\end{figure}

Figure \ref{fig:fig8} shows the mass accretion rate at the inner
radius $r_{\rm in}$ as a function of time.  The weak-field simulation
S2 shows no significant differences relative to the Bondi
solution; the factor of $\simeq 2$ decrease in $\dot M$ for
S2 in Figure \ref{fig:fig8} is due to the fact that thermal conduction
leads to small changes in the density and temperature profiles of the
flow.
In addition, there is no indication that the MTI is present, unlike in
the simulations at large radii discussed in the previous section. By
contrast, all of the simulations with $\beta \sim 1$ at $r_{\rm out}$
-- both with and without anisotropic conduction -- show a significant
decrease in $\dot M$ after a few free-fall times, with $\dot M \sim
0.01-0.03 \, \dot M_B$ at late times. We focus our interpretation on
the simulations with conduction, since our results for simulations
without conduction are similar to those presented in \citet{igu02}.
Note that the simulations with anisotropic conduction show episodic
``flares'' in $\dot M$ and other dynamical quantities that are not as
prominent in the MHD simulations; these are discussed more below.

\begin{figure}
\begin{center}
\includegraphics[scale=0.42]{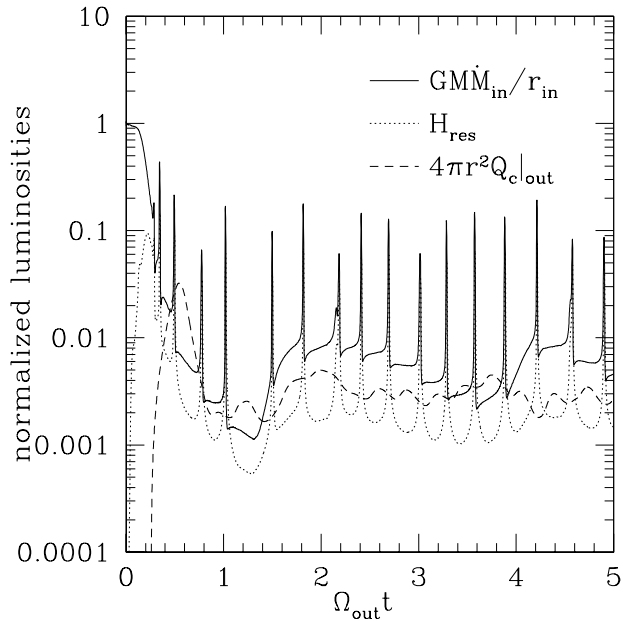}
\caption{Total energy generation and loss rates (normalized to
$GM\dot{M}_B/r_{\rm in}$) as a function of time for run S3 with
$\alpha_c=2$: gravitational energy released by accretion
$GM\dot{M}/r_{\rm in}$ (solid line), volume integrated resistive heating
rate ($H_{\rm}= 2 \pi \int dr d\theta r^2 \sin \theta \eta J^2$; dotted
line), and conductive luminosity at the outer radius of the domain
($4 \pi r^2Q_c|_{\rm out}$; short-dashed line).  Most of the energy
released by accretion is captured by our explicit resistivity; this
energy is then transported to large radii by conduction.}
\label{fig:enflow}
\end{center}
\end{figure}

Figure \ref{fig:enflow} shows the overall energetics for the
simulation with $\alpha_c = 2$ (S3).  A significant fraction of the
gravitational potential energy released by the inflowing matter (solid
line) is converted into magnetic energy by flux freezing.  Resistive
dissipation of oppositely directed field lines leads to plasma
heating.  Although our numerical routine is not conservative, we do
have an explicit resistivity and Figure \ref{fig:enflow} shows that
the total resistive heating in the computational domain (dotted line)
is similar to the gravitational potential energy released by
accretion.\footnote{In addition to explicit resistivity -- which is
the dominant source of heating in our calculations -- we also include
artificial viscosity for shock capturing.  Explicit resistivity,
artificial viscosity, and inherent numerical diffusion all lead to
increase in entropy in the simulations.}  In turn, much of this
energy is transported out of the computational domain by conduction,
as shown by the short-dashed line in Figure \ref{fig:enflow}, which is
the conductive luminosity at the {\it outer} radius $r_{\rm out}$.
The results for runs S1 \& S4 (which have different conductivities
$\alpha_c$) look similar.

\begin{figure}
\begin{center}
\includegraphics[scale=0.42]{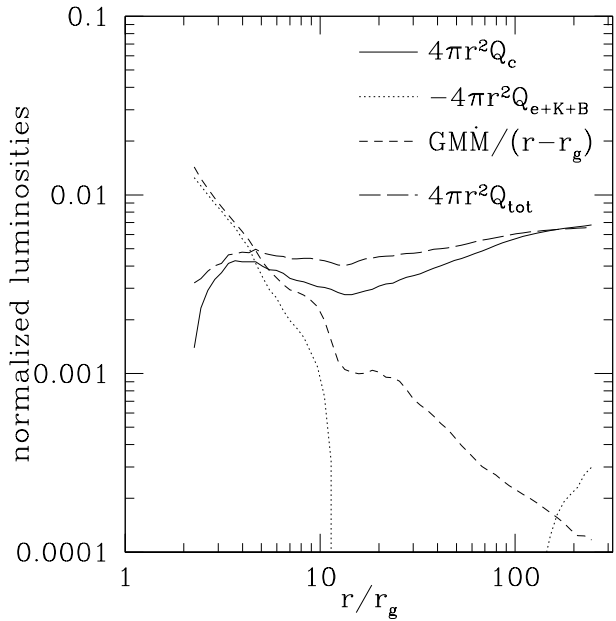}
\caption{Time and angle averaged luminosities ($4 \pi r^2 \times $
energy fluxes, normalized to $GM\dot{M}_B/r_{\rm in}$) as a function
of radius for run S3 ($\alpha_c = 2$): conductive luminosity $4 \pi
r^2 Q_c$, energy transport by fluid motion $4 \pi r^2Q_{e+K+B}$ (which
includes thermal, kinetic, and Poynting contributions), total power
generated by accretion ($GM\dot{M}/[r-r_g]$), and the total luminosity
($4 \pi r^2Q_{\rm tot}$, where $Q_{\rm
tot}=Q_c+Q_{e+K+B}+GM\dot{M}/4\pi[r-r_g]$).}
\label{fig:heatflux}
\end{center}
\end{figure}

Another indication of the importance of the conductive transport of
energy is given in Figure \ref{fig:heatflux}, which shows the time
averaged luminosities carried by conduction ($Q_c$) and fluid motion
($Q_{e+K+B}$) as a function of radius for $\alpha_c = 2$ (S3).  Note that
here we are comparing the conductive luminosity to the total 
thermal+kinetic+magnetic transport of energy 
$Q_{e+K+B} \equiv \rho V^2 V_r/2 + \gamma p
V_r/(\gamma-1) + [B^2V_r-({\bf B \cdot V})V_r]/4\pi$ which includes
kinetic, thermal, and Poynting contributions; the radial velocity
$V_r$ in $Q_{e+K+B}$ is the total radial velocity and thus $Q_{e+K+B}$ 
includes energy
transport by the mean fluid motion and by the turbulence (unlike $Q_t$
in \S 3 and Fig. \ref{fig:fig6} which is solely the turbulent
transport).  The energy carried by conduction is significantly larger
than that carried by other mechanisms (in particular, either advection
or convection) except at the smallest radii, where the
pseudo-Newtonian potential leads to supersonic inflow.  In addition,
the conductive luminosity is roughly independent of radius, as would
be expected in steady state given a source of heating at small
radii. These results demonstrate that the gravitational potential
energy released at small radii by magnetic dissipation is carried to
larger radii by conduction.  We find similar results for $\alpha_c =
0.2$ (S1) and $\alpha_c = 20$ (S4), but for larger values of
$\alpha_c$, the conductive energy transport becomes the dominant
energy transport mechanism at smaller radii. Only in the limit
$\alpha_c \ll 1$ do the simulations with conduction become similar to
the simulations of \citet{igu02}.

\begin{figure}
\begin{center}
\includegraphics[scale=0.42]{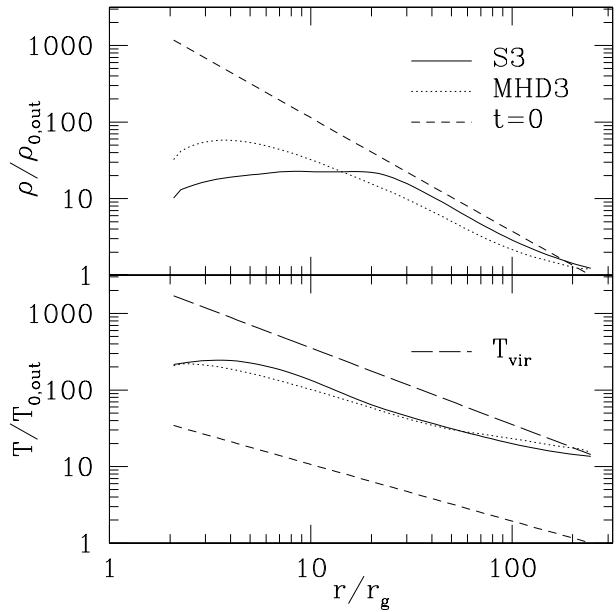}
\caption{Time and angle averaged density (upper panel) and temperature
(lower panel) as a function of radius for S3 ($\alpha_c = 2$) and MHD3
compared to the initial solutions; the profiles for S3 and MHD3 are
angle and time-averaged.  In both S3 \& MHD3, the plasma is heated to
nearly the virial temperature and the density in the inner part of the
flow decreases significantly. As shown in Fig. \ref{fig:2Din}, there
is a significant asymmetry between the polar and equatorial regions;
the density and temperature are larger at the equator while magnetic
pressure dominates at the poles. }
\label{fig:nT}
\end{center}
\end{figure}

\begin{figure}
\begin{center}
\includegraphics[scale=0.42]{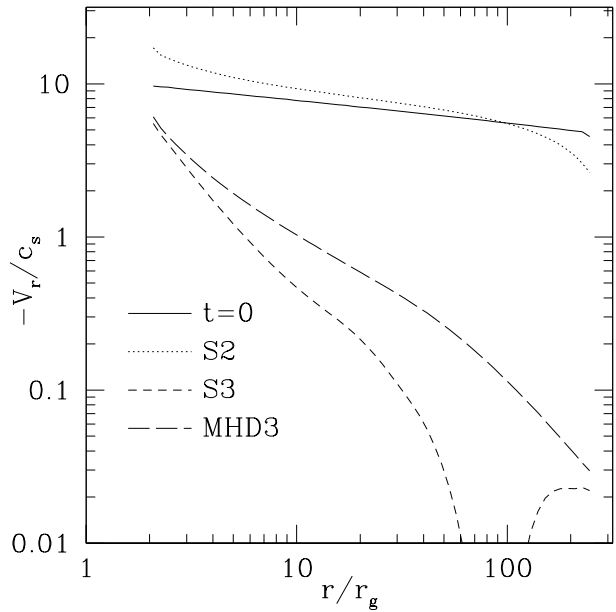}
\caption{The ratio of the time and angle averaged Mach number (radial
velocity divided by isothermal sound speed $c_s$) at late times as a
function of radius for the initial Bondi solution (solid line), S2
(dotted line), S3 (short dashed line), and MHD3 (long-dashed
line). For the calculations with significant magnetic dissipation (S3
\& MHD3), the average infall speed is subsonic at $r \gtrsim$ few
$r_g$, and the plasma is in rough hydrostatic equilibrium.
Although not shown, we find that the radial infall velocity in the
polar regions is larger than in the hotter equatorial regions.}
\label{fig:vr}
\end{center}
\end{figure}

Figure \ref{fig:nT} shows how the time and angle averaged density and
temperature profiles of the accretion flow change in the presence of
resistive heating and conduction.  The initial temperature of the flow
is much less than the virial temperature $kT_{\rm vir} \equiv GMm_p/r$
because we initialize a Bondi solution with $\gamma = 1.5 < 5/3$.  By
contrast, both the MHD simulation and the simulations with anisotropic
conduction have time-averaged temperatures of $T \sim T_{\rm vir}$
because much of the accretion energy is thermalized as heat via
resistive dissipation.  The resulting high temperatures lead to a flow
in approximate hydrostatic equilibrium, with a mean radial velocity
that is highly subsonic at $r \gtrsim \, {\rm a \, few} \, r_g$ (see
Fig. \ref{fig:vr}).  Figure \ref{fig:nT} also shows that the density
of the flow in the inner region is much lower in the simulations with
heating than in the initial solution.  This is consistent with
accretion rate being $\ll \dot M_B$ (Fig. \ref{fig:fig8}).

In Figures \ref{fig:heatflux}-\ref{fig:vr} we have presented
results averaging over the full range of $\theta$, from $\theta=0$ to
$\theta=\pi$. However, the initially vertical magnetic field lines
impose an asymmetry on the flow so that the flow profiles are not
spherically symmetric.  As seen in Figure \ref{fig:2Din} (discussed
below), the temperature is larger in the equatorial region compared to
the pole, by roughly a factor of 10 at small radii; the same is true
for the density.  The equatorial region is primarily gas pressure
dominated while the poles are magnetically dominated.  The relative
importance of the different energy transport mechanisms (shown in
Fig. \ref{fig:heatflux}) also varies with angle; in particular,
conduction dominates at the poles (and averaged over all angles), but
is less important in the equatorial region where the field lines are
less ordered.  Overall, the angular asymmetry we find in these
nonrotating simulations is not as large as in simulations with
significant angular momentum \citep[e.g.,][]{haw02} which show
``funnels" with radial outflows. In our spherical simulations, the
time averaged flow is always inwards at all angles (although there can
be outflow at certain times; see Fig. \ref{fig:2Din}).

Although there are modest quantitative differences between the
simulations with anisotropic conduction and those without (Figs.
\ref{fig:fig8}-\ref{fig:vr}), our results demonstrate that many of
the properties of magnetized spherical accretion are in fact rather
similar in the two cases.  An important difference, however, is that
for even modest thermal conductivities ($\alpha_c \sim 0.2$;
eq. [\ref{eq:cond}]), conduction is the dominant mechanism of energy
transport and transports a significant fraction of the energy released
by magnetic dissipation to large radii (Fig. \ref{fig:enflow} \&
\ref{fig:heatflux}).  By contrast, in the MHD simulations without
conduction, we find that the net energy transport is {\it inwards}.
Although convective motions are present, the convective luminosity is 
much smaller than $GM\dot{M}/(r-r_g)$ at all radii.

\begin{figure}
\begin{center}
\includegraphics[scale=0.42]{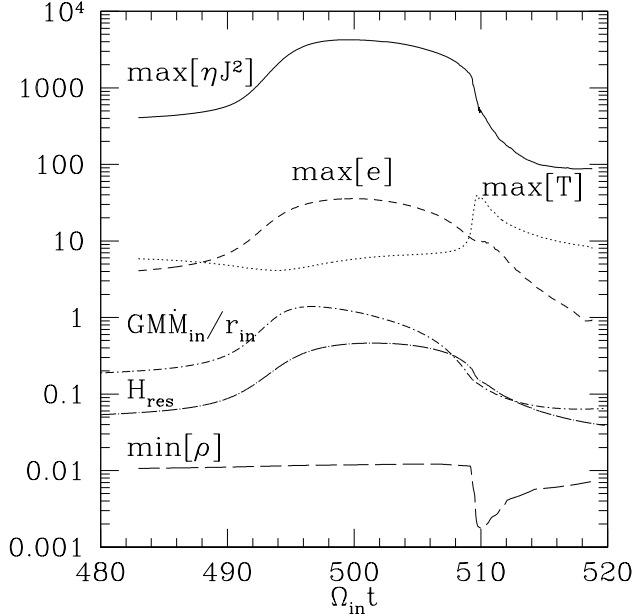}
\caption{Zoom-in on the time-dependence of an accretion/heating flare
for run S3; note that time here is in units of $\Omega_{\rm in}^{-1}
\equiv (G M/r_{\rm in}^3)^{-1/2}$ not $\Omega_{\rm out}^{-1}$.  The
maximum values of the resistive heating rate ($\eta J^2$), internal
energy per unit volume (e), and temperature (T), along with the
minimum value of the density ($\rho$), the energy released by
accretion ($GM\dot{M}_{in}/r_{\rm in}$), and the volume integrated
resistive heating rate ($H_{\rm res}$) are all shown as a function of
time (the max[X] and min[X] are taken over all grid points in the
computational domain).  All quantities are in arbitrary units.  As
shown in Figure \ref{fig:2Din}, the peaks in temperature, heating
rate, etc. occur at small radii near the equatorial plane.}
\label{fig:Tmax}
\end{center}
\end{figure}

\begin{figure}
\begin{center}
\includegraphics[scale=0.3]{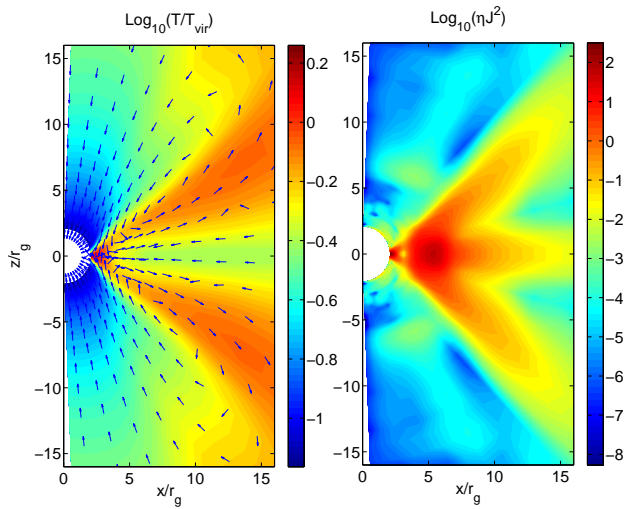}
\caption{ {\em Left:} Log$_{10}$ of plasma temperature normalized to
the virial temperature ($T/T_{\rm vir}$) in the inner 15 $r_g$ for run
S3 at $\Omega_{in}t=510$ (the peak in max[T] in
Fig. \ref{fig:Tmax}). Arrows show the direction of the {\it velocity}
vectors. {\em Right:} Log$_{10}$ of the local resistive heating rate,
$\eta J^2$ (in arbitrary units), at the same time. Both the resistive
heating and temperature peak at small radii in the equatorial region.}
\label{fig:2Din}
\end{center}
\end{figure}

An additional interesting difference between the simulations with and
without conduction is the ``flaring'' in the simulations with
anisotropic conduction (Fig. \ref{fig:fig8}-\ref{fig:enflow}).  The
accretion rate, resistive heating, and conduction luminosity all show
regular rapid increases (``flares'') by a factor of $\sim 10-100$ on a
timescale $\sim 10$ orbital periods at $\sim r_{\rm in}$ (see Fig. 
\ref{fig:Tmax}).  The
characteristic timescale between such flares is a fraction of an
orbital period at $r_{\rm out}$.  The flare timescales do not depend
sensitively on the value of the thermal conductivity, as long as 
$\alpha \simeq 2$. As shown in Fig. \ref{fig:fig8}, flares are most 
prominent for run S2, the run with the smallest mass accretion rate 
among $r \ll r_B$ simulations (see Table \ref{tab:tab2}).

To assess the robustness of this flaring, we performed several checks.
Similar flaring is much less pronounced in the MHD simulations and in
simulations with an isotropic thermal conductivity, even though the
reduction in $\dot M$ relative to $\dot M_B$ is similar; in addition,
the flaring is more pronounced for larger values of $\alpha_c$.  To
assess whether the flaring is a numerical artifact, we carried out
simulations implementing anisotropic conduction using the method of
\citet{par05} and \citet{sha06b}, which uses simple finite
differencing (our method, based on \citealt{sha07c}, limits the heat
fluxes so that the temperature is always positive).  We find quite
similar flaring with the two methods.\footnote{At late times in the 
simulations
with conduction implemented using simple finite differencing, the
temperature can become negative, which demonstrates the importance of
using a method that guarantees that heat flows from hot to cold, such
as that of \citet{sha07c}.} We also changed the magnitude of
the artificial viscosity and the explicit resistivity, and the form
of the explicit resistivity (e.g., $\eta \propto dr
(c_s^2+V_A^2)^{1/2}$ instead of eq.  \ref{res}, where $c_s$ and $V_A$
are the local isothermal sound speed and Alfv\'en speed, respectively),
but the flaring in run S3 (Fig.  \ref{fig:fig8}) persisted.  These
checks argue against the flaring being a numerical artifact.  The
physical origin of this flaring is, however, not clear.  The MTI could
in principle be present because the inflow is subsonic, but it should
be stabilized by magnetic tension.  The flaring we find is
qualitatively similar to the ``relaxation oscillations'' seen in some
1D time-dependent models of spherical accretion with radiative heating
\citep[e.g.,][]{cow78} and could be analogous, with spatially
extended heating and redistribution of energy by anisotropic
conduction precluding the existence of a steady state solution.

Figure \ref{fig:Tmax} shows a number of quantities characterizing one
of the accretion/heating flares over a small interval in time.  The
gravitational potential energy released by accretion ($GM\dot{M}_{\rm
in}/r_{\rm in}$) peaks slightly earlier in time than the volume
integrated resistive heating rate ($H_{\rm res}$); this is consistent
with the field being amplified by flux freezing and then dissipated by
resistive diffusion.  Figure \ref{fig:Tmax} also shows the maximum
values of the local resistive heating rate ($\eta J^2$), the internal
energy per unit volume ($e$), and the temperature, and the minimum
value of the density (where the maximum/minimum are over all grid
cells in the computational domain).  The maximum resistive heating
rate and maximum internal energy have a time-dependence similar to
that of the accretion energy.  The peak temperature and minimum
density, on the other hand, occur after the majority of the resistive
heating has taken place, and their evolution in time is much more
rapid.  In fact, the increase in temperature (decrease in density)
occurs on a timescale comparable to the free-fall timescale at small
radii.

Figure \ref{fig:2Din} shows contour plots of temperature relative to
the virial temperature and local resistive heating rate (in arbitrary
units) in the inner $15 r_g$ at the time of the peak in max[T] in
Figure \ref{fig:Tmax} ($\Omega_{\rm in} t \simeq 510$).  Also shown
are the {\it velocity} unit vectors.  The resistive heating is
strongly concentrated in the equatorial region at $\lesssim 5 r_g$
(near the sonic point; see Fig. [\ref{fig:vr}]) where rapid inflow
brings together oppositely directed field lines.  Strong heating leads
to modestly supervirial temperatures at small radii.  This energy
diffuses to large radii along a finite set of field lines because of
the anisotropic nature of conduction.  The supervirial temperatures
also reverse the inflow along particular field lines, leading to a 
weak mass outflow 
(as shown by the velocity unit vectors in Fig. \ref{fig:2Din}).

\section{Discussion}

We have studied the effects of anisotropic thermal conduction on the
dynamics of spherical accretion flows using global axisymmetric
simulations.  Our simulations address two different physical effects
that operate on different radial scales in spherical accretion flows.
\subsection{The MTI in Spherical Accretion Flows}
We have calculated the nonlinear saturation of the magnetothermal
instability (MTI) at radii $\sim r_B$ in spherical accretion flows (\S
3), where $r_B \simeq GM/c_s^2$ is the gravitational sphere of
influence of the central object (of mass $M$) and $c_s$ is the isothermal 
sound speed of the ambient plasma.  The MTI does not grow appreciably at
radii $\ll r_B$ for the canonical Bondi spherical accretion solution
because the inflow is supersonic.  Our simulations are the first
global simulations of the MTI, but our results are largely consistent
with the previous local simulations of \citet{par05,par07}.  We have
focused on studying the saturation of the MTI for initially weak
vertical magnetic fields ($\beta \gg 1$).

We find that the MTI saturates by aligning the magnetic field lines
with the temperature gradient, i.e., by making the field lines
primarily radial (Figs. \ref{fig:fig2} - \ref{fig:fig3b}). This occurs
on approximately the local dynamical time ($\sim [g|d\ln
T/dz|]^{-1/2}$).  Since the field lines end up largely radial, the
rearrangement of the magnetic field results in a radial heat flux
similar to the field-free (Spitzer) value; in the saturated state,
energy transport by conduction dominates transport by the convective
motions associated with the MTI (Fig. \ref{fig:fig6}).  Given the
significant radial heat flux, the density and temperature profiles of
the accretion flow adjust to satisfy ${\bf \nabla \cdot Q} \simeq
0$, where ${\bf Q}$ is the conductive flux of energy.  In all cases
of $\kappa(r)$ that we considered (see \S \ref{sec:kappa}), the mass
accretion rate in the presence of the MTI differs only modestly from
the canonical Bondi rate (Fig. \ref{fig:fig5}).

Our results indicate that the saturation of the MTI is quasilinear in
the weak field limit: the magnetic and convective energy densities in
the saturated state are roughly proportional to the initial magnetic
energy density (Fig. \ref{fig:fig7}).\footnote{Ian Parrish has found
this same result in local three-dimensional Cartesian simulations
(private communication).}  The net result is that the MTI amplifies
the magnetic field by a factor of $\sim 10-30$ {\it independent of the
initial magnetic field strength} (for weak fields).  For stronger
initial fields, magnetic tension can saturate the MTI before the field
has become largely radial \citep{par05,par07}.  In this case, the
amplification of the magnetic field will be less than we have found
here.

The quasilinear saturation of the MTI can be understood qualitatively
using linear theory \citep{qua08}; as a result, we believe that this
is a general property of the MTI, rather than being specific to the
spherical accretion problem considered in this paper. The linear
growth rate of the MTI is reduced by a factor of $\sim B_\theta/B$ for
nearly radial fields, and the growth occurs on progressively smaller
scales for small $B_\theta/B$ (eq. [\ref{eq:growth}]).  Thus, as the
field lines become radial, the MTI effectively shuts off its own
growth.  This is in stark contrast to the magnetorotational
instability in differentially rotating plasmas, which is an exact
nonlinear solution of incompressible MHD, and for which growth
continues unimpeded even when $\delta B \gg B$ \citep{goodman94}.

One simplification in the current calculations is that we have
not included anisotropic viscosity; the diffusion coefficient for
anisotropic viscosity in a collisional plasma is smaller than the
diffusion coefficient for anisotropic conduction by a factor of $\sim
(m_p/m_e)^{1/2} \simeq 40$; in a collisionless plasma, the relative
importance of these two effects is more complex and depends on
small-scale kinetic microinstabilities (e.g., Sharma et al. 2007).
Local simulations of the MRI find that the saturation of the
instability is sensitive to the ratio of the resistivity to the
viscosity (e.g., Fromang et al. 2007).  In future calculations, the
effects of anisotropic viscosity on the evolution of the MTI should
thus be studied, but this is probably first best done in a local
calculation, rather than in the global geometry considered here.

\subsubsection{Application to the Galactic Center}

One potential application of the MTI at large radii in spherical
accretion flows is to the observed Faraday rotation from the Galactic
center.  Millimeter observations of Sgr A* find a rotation measure
(RM) of $\simeq - 6 \times 10^5$ rad m$^{-2}$ that has been roughly
constant over the past $\sim 8$ years \citep{ait00,bow05,mar07}.
Propagation through the interstellar medium cannot account for this
large RM.  In addition, X-ray observations of the thermal plasma in
the central parsec of the Galactic center find that the electron
number density is $\approx 100$ cm$^{-3}$ and the temperature is
$\approx 2$ keV at a distance of $\approx 10^5 r_g$ from the black
hole \citep{bag03,qua04}.  If the magnetic field on this scale were
uniform, radial, and in rough equipartition with the thermal pressure,
$B \approx 2 \,\beta^{-1/2}$ mG and $RM \approx 5000 \, \beta^{-1/2}$
rad m$^{-2}$, two orders of magnitude smaller than what is observed.
This shows that the observed RM must be produced at distances $\ll
10^5 r_g$ from Sgr A*.

Our calculations show that given a sufficiently coherent seed magnetic
field at large radii, the MTI will generate a relatively coherent
radial field at distances of $\sim r_B \sim 10^5 r_g$ from Sgr A*.
Given the measured density and temperature of the plasma, the electron
mean free path is $\sim r_B$ at radii $\sim r_B$ and thus the MTI
grows on a dynamical time.  Numerical simulations of accretion onto
Sgr A* from stellar winds located at $\sim 10^5-10^6 r_g$ find that
the circularization radius of the flow is $r_{\rm circ} \sim 10^3-10^4
r_g$ \citep{cua06}.  If the Bondi solution is applicable from the
scale of the observed plasma at $\sim r_B$ to $\sim r_{\rm circ}$,
then $d RM/d \ln r \propto r^{-7/4}$ at fixed $\beta$ and the RM
produced by the roughly spherical inflowing plasma at large radii
around Sgr A* is \be RM \sim 10^7 \, \beta^{-1/2} \, \left(r_{\rm
circ} \over 10^3 r_g\right)^{-7/4} \, \rad. \label{eq:RM}\ee It is
difficult to determine with confidence whether the stellar winds
feeding Sgr A* -- or the shocks between such winds -- produce a
sufficiently large and coherent 'seed' magnetic field at $\sim r_B$ to
account for the RM towards Sgr A*.  However, for an initial value of
$\beta$ of $\beta_0$ (assumed $\gg 1$) at $r_B \sim 10^5 r_g$, the
combined action of the MTI and flux freezing will decrease $\beta$ to
$\sim 10^{-6} \beta_0 (r_{\rm circ}/10^3 \, r_g)^{-3/2}$ at $r_{\rm
circ}$.  Thus even a relatively weak magnetic field ($\beta_0 \sim
10^6-10^8$; $B_0 \sim 0.1-1 \mu G$) on scales of $\sim 10^5 r_g \sim
0.1$ pc will be amplified by the MTI and flux freezing to a value
capable of explaining the observed RM from Sgr A*.\footnote{If most O
stars have surface magnetic fields comparable to the $\sim$ kG field
measured in the young O star $\theta^1$ Orinis C \citep{donati2002},
stellar winds alone will fill the central $\sim 0.1$ pc with a
magnetic field $\sim$ mG.}  Our explanation for the observed RM also
naturally accounts for its stability, since the dynamical time at
$\sim 10^3-10^4 r_g$ is $\sim 0.3-10$ years.  By contrast, if the RM
is produced in the rotating accretion flow even closer to the black
hole, it is harder to account for the observed stability
\citep{sha07b}.

\subsection{The Effects of Thermal Conduction on the Dynamics of Spherical Accretion}

The second set of simulations we have carried out address the effects
of anisotropic conduction and resistive dissipation of magnetic energy
on the dynamics of spherical accretion flows at small radii $\sim
2-256 \, r_g$ (\S 4).  Our results are in reasonable agreement with those
of \citet{igu02}, who studied the same physical problem but without
anisotropic conduction.  Although the MTI does not dramatically modify
the dynamics of spherical accretion at $\sim r_B$, it does generate a
significant radial magnetic field that can be further amplified by
flux freezing.  For $r \ll r_B$, supersonic inflow in
(magneto)hydrodynamic Bondi accretion leads to magnetic field
amplification by flux freezing, converging field lines, and resistive
heating of the plasma.  When the magnetic energy density is comparable
to the gravitational energy density, resistive heating can heat the
plasma up to the virial temperature and significantly modify the
dynamics of spherical accretion \citep{igu02}.  This is only likely to
occur for sub-Eddington accretion rates onto compact objects, since
only under these conditions is radiative cooling negligible.

All of our calculations with resistivity and strong magnetic fields at
$\sim 100 \, r_g$ -- both with and without anisotropic conduction -- show
a significant decrease in $\dot M$ after a few free-fall times, with
$\dot M \sim 0.01 \dot M_B$ in the quasi steady state at late times
(Fig. \ref{fig:fig8}).  The high temperatures generated by resistive
heating stifle the inflow, reducing the accretion rate
(Fig. \ref{fig:fig8}) and generating a subsonic (Fig. \ref{fig:vr})
nearly hydrostatic atmosphere around the central object.  For
realistic values of the thermal conductivity for collisionless plasmas
(see \S \ref{sec:kappa}), we find that conduction transports the
majority of the energy released by accretion to large radii
(Figs. \ref{fig:enflow} \& \ref{fig:heatflux}).  By contrast, although
turbulent motions are present, there is no net convective energy flux
from small to large radii, even in our simulations {\it without}
conduction (Fig. \ref{fig:heatflux}).\footnote{Our interpretation of
the MHD simulations of spherical accretion without conduction thus
differs 
from that of \citet{igu02}, who emphasized the importance of radial
energy transport by convection.  This needs to be investigated
further.}  This indicates that significant radial transport of energy
(be it conductive or convective) is not necessary to suppress
accretion onto the central object; instead, heating that raises the
temperature to the virial temperature appears sufficient.  Indeed, it
is well-known that there is no supersonic Bondi solution for adiabatic
indices $\gamma > 5/3$. Heating via resistive dissipation and/or
artificial/numerical viscosity effectively increases the adiabatic
index above this value, thus stifling accretion onto the central
object. Thermal conduction can further reduce the accretion rate by
conducting the dissipated energy to larger radii.

Overall, our results are in reasonable agreement with \citet{joh06}'s
simplified one-dimensional models for the structure of spherical
accretion in the presence of conduction and plasma heating.  Our
simulations confirm their hypothesis that a significant fraction of
the energy released by accretion is transported to large radii by
thermal conduction.  In addition, their prediction that $\dot M/\dot
M_B$ should decrease by $\simeq 2-3$ orders of magnitude -- with a
factor of $\sim 10$ suppression due to heating alone (their
Fig. 8) -- is consistent with our numerical results.

The reduction in $\dot M$ that we find is similar to that needed to
explain the low luminosity of Sgr A* (e.g., \citealt{sha07b}).  It is
also reasonably consistent with the limits on $\dot M$ from Faraday
rotation measurements (e.g., \citealt{mar07}).  However, it is
unlikely that the present calculations are quantitatively applicable
to Sgr A* given the estimated circularization radius of the accretion
flow in Sgr A* of $r_{\rm circ} \sim 10^3-10^4 r_g$ \citep{cua06}.
The gravitational potential energy released accreting to $\sim r_{\rm
circ}$ is unlikely to significantly modify the accretion flow
dynamics.  Thus the accretion rate and outward conductive energy flux
will be largely determined in the rotationally supported flow at radii
$\lesssim r_{\rm circ}$.  Nonetheless, our results demonstrate that
even in the absence of angular momentum, the Bondi accretion solution
does not apply to dilute magnetized accretion flows.  Instead, the
accretion flow is heated to nearly the virial temperature by resistive
dissipation of magnetic energy, the accretion rate is $\ll \dot M_B$,
and most of the gravitational potential energy released by accretion
is transported to large radii by thermal conduction.  In addition, in
our simulations with anisotropic conduction, the accretion and heating
are time-dependent, with episodic ``flaring''
(Figs. \ref{fig:fig8}-\ref{fig:enflow}) followed by mass outflow along
a subset of the field lines (Fig. \ref{fig:2Din}).  It is important to
determine if analogous effects are present in rotating accretion flows
(which is by no means clear), given the possible application to the
broad-band electromagnetic flares observed from Sgr A*
\citep[e.g.,][]{mar08} and other systems.

\subsection{Additional Applications}

In addition to the applications considered here, thermal conduction is
also important in the intracluster medium of galaxy clusters
\citep[][]{sar86}.  A conductive energy flux from the accretion flow
onto the central black hole in the nucleus of galaxy clusters could
help heat the intercluster plasma at radii $\sim 0.1-1$ kpc and
prevent it from cooling catastrophically.  At larger radii, the
temperature in clusters increases from $r_B \sim 0.1$ kpc to $\sim
100$ kpc, where it begins to decrease \citep{piffaretti05}.  Thus the
plasma at intermediate radii in clusters appears stable to the MTI (it
is, however, unstable to the HBI; \citealt{qua08,par08}).  However,
\citet{cha06} have shown that cosmic rays from a central active
galactic nucleus (AGN) can make the plasma in the cores of clusters
unstable to a cosmic-ray mediated version of the MTI if $dp_{CR}/dr +
nk dT/dr < 0$ ($p_{CR}$ is the cosmic ray pressure), i.e., if there is
a sufficiently large cosmic ray flux from the central AGN. It is
likely that the MTI in the presence of cosmic rays saturates in a
manner analogous to that found here, by aligning the field lines
primarily in the radial direction, with a roughly constant radial
cosmic ray and conductive luminosity.  One of the outcomes of this
work (see also \citealt{par07}) is that convective transport of energy
is subdominant in the presence of the MTI; this may rule out turbulent
heating due to cosmic-ray driven convection \citep[][]{cha07} as an
important source of plasma heating in cluster cores.  However, if the
cosmic-ray mediated MTI generates a largely radial magnetic field in
cluster cores, the conductive energy flux from large radii can produce
significant heating at small radii, analogous to what we have found in
our global simulations (see Fig. [\ref{fig:fig6}]).  It is clear that
a careful study of the combined action of the MTI and HBI in the
presence of cosmic-rays is necessary to understand the thermal
structure of galaxy clusters.

\section*{Acknowledgments}

We thank Bryan Johnson and Ian Parrish for useful discussions.  We thank 
the referee Steve Balbus for detailed comments. PS and
EQ were supported in part by NASA grant NNG06GI68G and the David and
Lucile Packard Foundation. PS was partly supported by DOE award
DE-FC02-06ER41453 to Jon Arons. JS was supported by DOE grant
DE-FG52-06NA26217. This research used resources of the National Energy
Research Scientific Computing Center, which is supported by the Office
of Science of the U.S. Department of Energy under Contract
No. DE-AC02-05CH11231. Computational resources were also provided by
the Princeton Plasma Physics Laboratory Scientific Computing Cluster.
Part of the code was tested with the Tungsten cluster at the National 
Center for Supercomputing Applications at the University of Illinois at 
Urbana-Champaign.

\label{lastpage}

\end{document}